# Influence of Induced Charges in the Electric Aharonov-Bohm Effect

Rui-Feng Wang

Department of Physics,　Beijing Jiaotong University,

Beijing　100044　China

E-mail: aboud56789@163.com

This paper states that the induced charge should not be neglected in the electric Aharonov-Bohm effect. If the induced charge is taken into account, the interference pattern of the moving charge will not change with the potential difference between the two metal tubes. It means that the scale potential itself can not affect the phase of the moving charge, and the true factor affecting the phase of the moving charge is the energy of the system including the moving charge and the induced charge.

In classical physics, the concept of "force" is the most important, all the phenomena can be explained by the forces acting on the objects. In classical electrodynamics, the Lorentz force acting on a charge is determined by the electric field **E** and the magnetic field **B** at the position of the charge. So, the electric field **E** and the magnetic field **B** are considered as more fundamental quantities than the scale potential $\varphi$ and the vector potential **A**. In quantum mechanics, what appear in the Schrödinger equation are the scale potential $\varphi$ and the vector potential **A** instead of **E** and **B**. So, some physicists asserted that the potential functions $\varphi$ and **A** are fundamental than **E** and **B**[1]. Just for this reason, Y. Aharonov and D. Bohm predicted a new effect named by their names later [2]. This new effect asserts that the phase of a moving charge will be changed by the potential functions $\varphi$ and **A**, even though the charge always move in a region where both **E** and **B** are zero, but the $\varphi$ and **A** are not zero. This effect includes the electric A-B effect and the magnetic A-B effect. The magnetic A-B effect has been studied extensively in both theory and experiments [3-11]. The existence of the magnetic A-B effect has been supported by some experiments [3,5,9]. But the electric Aharonov-Bohm effect was much less studied [12,13,14].

This paper will focus on the electric A-B effect in theory, especially, the possible experiment with two metal tubes proposed in the seminal paper by Aharonov and Bohm. It is found that a very important factor was neglected in their paper [2]. That is the induced charge on the inner surfaces of the metal tubes. If this induced charge is taken in to account, we will find the phase of the moving charge will not change with the electric potential difference between two metal tubes. So, it is more convincing to state that the real factor influencing the phase of a moving charge is the energy of the system including the moving charge and the induced charge, but not the scale potential itself. The similar conclusion about the magnetic A-B effect [6,10,11] and its theoretical proof [11] have also been proposed.

First let us repeat the possible experiment proposed by Aharonov and Bohm to demonstrate the electric A-B effect. As depictured in Fig.1, a coherent charge beam is split into two parts and each part enters into a separate long cylindrical metal tube. After the beams pass through the metal tubes, they are combined to interfere coherently at F. By means of time-determining electrical "shutters" the charge beam is divided into wave packets, the length of each wave packet is long compared with its wavelength but short compared with the length of the metal tube. To analysis this experiment in more detail, we suppose the moving charge enter into the metal tubes at $t_0$, comes out from the tubes at $t_3$, in addition, $t_0 < t_1 < t_2 < t_3$. During the time interval from $t_1$ to $t_2$, while the moving electron is well inside the tubes, a potential difference $V_0$ is applied between these two tubes. For example, the tube 1 is always connected to the zero potential point, and the tube 2 is connected to a external voltage generator, which makes the electric potential of the tube 2 to alternate in time as following:

$$U_2(t) = \begin{cases} 0 & t < t_1 \\ V_0 & t_1 \leq t \leq t_2 \\ 0 & t > t_2 \end{cases} \quad (1)$$

To keep the potentials of the two tubes zero in $t < t_1$ and $t > t_2$, the metal tube 2 should also be connected with the zero potential point in these two time intervals. Otherwise, the collision of the ions and the charges accumulation will make the potential of the tube 2 uncontrollable.

Let's discuss this problem in the following situations:

*The first situation*: The external voltage generator is switched off, $\psi_1^0(x,t)$

and $\psi_2^0(x,t)$ represent the wave functions of the parts passing through the tubes 1 and 2, respectively, which are unperturbed by the external electric potential; $x$ ($x = x_1, x_2, x_3$) is the coordinate of the moving charge. The total wave function $\psi^0(x,t)$ is:

$$\psi^0(x,t) = \psi_1^0(x,t) + \psi_2^0(x,t) \tag{2}$$

$\psi_1^0(x,t)$ and $\psi_2^0(x,t)$ are determined by the following equations:

$$\begin{cases} i\hbar \dfrac{\partial}{\partial t}\psi_1^0(x,t) = -\dfrac{\hbar^2}{2m}\dfrac{\partial^2}{\partial x^2}\psi_1^0(x,t) \\ i\hbar \dfrac{\partial}{\partial t}\psi_2^0(x,t) = -\dfrac{\hbar^2}{2m}\dfrac{\partial^2}{\partial x^2}\psi_2^0(x,t) \end{cases} \tag{3}$$

where $\dfrac{\partial^2}{\partial x^2} \equiv \dfrac{\partial^2}{\partial x_1^2} + \dfrac{\partial^2}{\partial x_2^2} + \dfrac{\partial^2}{\partial x_3^2}$.

*The second situation*: The external voltage generator is switched on, $\psi_1(x,t)$ and $\psi_2(x,t)$ represent the wave functions perturbed by the external electric potential. The total wave function $\psi(x,t)$ is:

$$\psi(x,t) = \psi_1(x,t) + \psi_2(x,t) \tag{4}$$

$\psi_1(x,t)$ and $\psi_2(x,t)$ are determined by the following equations:

$$\begin{cases} i\hbar \dfrac{\partial}{\partial t}\psi_1(x,t) = -\dfrac{\hbar^2}{2m}\dfrac{\partial^2}{\partial x^2}\psi_1(x,t) \\ i\hbar \dfrac{\partial}{\partial t}\psi_2(x,t) = -\dfrac{\hbar^2}{2m}\dfrac{\partial^2}{\partial x^2}\psi_2(x,t) + qU(t)\psi_2(x,t) \end{cases} \tag{5}$$

Then, comparing the eq.(3) and (5), the wave functions $\psi_1(x,t)$ and $\psi_2(x,t)$ for the charge in the two beams are given by:

$$\begin{cases} \psi_1(x,t) = \psi_1^0(x,t) \\ \psi_2(x,t) = \psi_2^0(x,t)\exp\left(\dfrac{-iq}{\hbar}\int_0^t U(t')dt'\right) \end{cases} \tag{6}$$

When $t > t_2$, the total wave function become:

$$\psi(x,t) = \psi_1^0(x,t) + \psi_2^0(x,t)\exp\left[\dfrac{-iqV_0(t_2 - t_1)}{\hbar}\right] \quad (t > t_2) \tag{7}$$

Comparing the equations (2) and (7), after the two beams come out from the tubes, an addition phase difference between these two beams appear in the second situation, which is

$$\Delta\varphi = \varphi_1 - \varphi_2 = \frac{qV_0(t_2 - t_1)}{\hbar} \tag{8}$$

So, when these two beams meet at F, the interference pattern will change with the value of $qV_0(t_2 - t_1)$.

The discussion above is quite similar to the original paper by Aharonov and Bohm[2], which also appears in many modern quantum mechanics textbooks[15]. But, in the discussion above, no induced charge was taken into account. No induced charge appearing on the surfaces of the metal tubes means that the electric field of the moving charge exists in all the space. So the electric field of the moving charge is not zero in the interior of the metal in these two situations above, furthermore, this electric field can penetrate the metal tube and exists in the region outside the metal tubes. Obviously, the discussion above is not reliable in principle. The induced charges should be taken into account.

In fact, once the moving charge $q$ (suppose $q = \pm e$) get into the metal tubes, an induced charge $q'$ ($q' = -q$) will appear on the inner surfaces of the metal tubes. ( If the moving charge $q$ is a proton, the induced charge $q'$ will be an electron; If the moving charge $q$ is a electron, then the induced charge $q'$ will be a hole in the Fermi gas of the metal tube, which is located on the inner surfaces of the tubes.) At the same time, another induced charge $q''$ ($q'' = q$) will appear on the outer surfaces of the metal tubes. During the time interval from $t_0$ to $t_1$, both the tubes are connected to the zero potential point, so, the induced charge $q''$ on the outer surfaces of the metal tubes will flow into the zero potential point. Therefore, only the induced charge $q'$ on the inner surfaces of the tubes needed to be taken into account.

For the induced charge $q'$ appears on the inner surfaces of the tubes, the electric field of the moving charge $q$ will be shielded by the induced charge $q'$, and the total electric field produced by the charges $q$ and $q'$ only exists in the region enclosed by the inner surface of the metal tubes. For the moving charge $q$ and the induced charge $q'$ attract each other, the distribution of the induced charge $q'$ changes with the position of the moving charge $q$, at the same time, the wave function of the moving charge $q$ is also perturbed by the induced charge $q'$[16, 17]. So, the moving charge $q$ is not a free particle in the metal tubes, we should take the moving charge $q$ and the induced charge $q'$ as a system. Let $\psi^0(x,y,t)$ and $\psi(x,y,t)$ represent the wave functions of this system with the external voltage generator being switched off or on, respectively, where, $x$ ($x = x_1, x_2, x_3$) denotes the coordinate of the moving particle $q$, and $y$ ($y = y_1, y_2, y_3$) denotes the coordinate of the induced particle $q'$.

In the region outside the metal tube, the moving charge is a free particle and has no relationship with the induced charges and the potential difference $U(t)$ between the two tubes. So, we need only to discuss the revolution of the wave function of the charges $q$ and $q'$ in the region enclosed by the inner surfaces of the metal tubes, *i.e.* the time dependence of the wave function from $t_0$ to $t_3$.

So, *the third situation*: the external voltage generator is switched off, but, the induced charge $q'$ is included. Then the total wave function is:

$$\psi^0(x,y,t) = \psi_1^0(x,y,t) + \psi_2^0(x,y,t) \qquad t_0 < t < t_3 \qquad (9)$$

where $\psi_1^0(x,y,t)$ and $\psi_2^0(x,y,t)$ represent the wave functions of the parts passing through the tubes 1 and 2, respectively, which are unperturbed by the external electric potential $U(t)$. The wave functions $\psi_1^0(x,y,t)$ and $\psi_2^0(x,y,t)$ satisfy the following Schrödinger equations :

$$\begin{cases} i\hbar \dfrac{\partial}{\partial t}\psi_1^0(x,y,t) = -\dfrac{\hbar^2}{2m}\dfrac{\partial^2}{\partial x^2}\psi_1^0(x,y,t) + \left\{ -\dfrac{\hbar^2}{2m'}\dfrac{\partial^2}{\partial y^2}\psi_1^0(x,y,t) + \phi(y)\psi_1^0(x,y,t) \right\} \\ \qquad\qquad + \dfrac{qq'}{4\pi\varepsilon_0 |x-y|^2}\psi_1^0(x,y,t) \hfill (10.a) \\[2ex] i\hbar \dfrac{\partial}{\partial t}\psi_2^0(x,y,t) = -\dfrac{\hbar^2}{2m}\dfrac{\partial^2}{\partial x^2}\psi_2^0(x,y,t) + \left\{ -\dfrac{\hbar^2}{2m'}\dfrac{\partial^2}{\partial y^2}\psi_2^0(x,y,t) + \phi(y)\psi_2^0(x,y,t) \right\} \\ \qquad\qquad + \dfrac{qq'}{4\pi\varepsilon_0 |x-y|^2}\psi_2^0(x,y,t) \hfill (10.b) \end{cases}$$

where, $t_0 < t < t_3$; $m'$ is the mass of the induced charge $q'$; $\phi(y)$ is the potential function experienced by the induced charge $q'$ in the metal tubes, which ensure the induced charge $q'$ can only move in the interior of the metal and can not leave out from the surfaces of the matel tubes, (noticing: $\phi(y)$ is independent with the potential difference $U(t)$ between the two matel tubes.) ;

$|x-y|^2 \equiv (x_1-y_1)^2 + (x_2-y_2)^2 + (x_3-y_3)^2$; and $\dfrac{qq'}{4\pi\varepsilon_0 |x-y|^2}$ is the interaction energy between the moving charge $q$ and the induced charge $q'$.

*The fourth situation*: the external voltage generator is switched on, and the perturbed wave function $\psi(x,y,t)$ of the system will become

$$\psi(x,y,t) = \psi_1(x,y,t) + \psi_2(x,y,t) \qquad t_0 < t < t_3 \qquad (11)$$

where $\psi_1(x,y,t)$ and $\psi_2(x,y,t)$ represent the wave functions perturbed by the external electric potential. $\psi_1(x,y,t)$ and $\psi_2(x,y,t)$ satisfy the following schrödinger equations:

$$\begin{cases} i\hbar\dfrac{\partial}{\partial t}\psi_1(x,y,t) = -\dfrac{\hbar^2}{2m}\dfrac{\partial^2}{\partial x^2}\psi_1(x,y,t) + \left\{-\dfrac{\hbar^2}{2m'}\dfrac{\partial^2}{\partial y^2}\psi_1(x,y,t) + \phi(y)\psi_1(x,y,t)\right\} \\ \qquad\qquad + \dfrac{qq'}{4\pi\varepsilon_0|x-y|^2}\psi_1(x,y,t) \qquad\qquad\qquad\qquad\qquad (12.a) \\ i\hbar\dfrac{\partial}{\partial t}\psi_2(x,y,t) = -\dfrac{\hbar^2}{2m}\dfrac{\partial^2}{\partial x^2}\psi_2(x,y,t) + \left\{-\dfrac{\hbar^2}{2m'}\dfrac{\partial^2}{\partial y^2}\psi_2(x,y,t) + \phi(y)\psi_2(x,y,t)\right\} \\ \qquad\qquad + \dfrac{qq'}{4\pi\varepsilon_0|x-y|^2}\psi_2(x,y,t) + qU(t)\psi_2(x,y,t) + q'U(t)\psi_2(x,y,t) \quad (12.b) \end{cases}$$

In the (12.b), $qU(t)$ and $q'U(t)$ are the potential energies of the moving charge $q$ and the induced charge $q'$ in the electric field $U(t)$, respectively. For $q=-q'$, the sum of $qU(t)\psi_2(x,y,t)$ and $q'U(t)\psi_2(x,y,t)$ is zero i.e. the interaction energy between $q$ and $U(t)$ is completely counteracted by the interaction energy between $q'$ and $U(t)$. Therefore, the equation (12) will become:

$$\begin{cases} i\hbar\dfrac{\partial}{\partial t}\psi_1(x,y,t) = -\dfrac{\hbar^2}{2m}\dfrac{\partial^2}{\partial x^2}\psi_1(x,y,t) + \left\{-\dfrac{\hbar^2}{2m'}\dfrac{\partial^2}{\partial y^2}\psi_1(x,y,t) + \phi(y)\psi_1(x,y,t)\right\} \\ \qquad\qquad + \dfrac{qq'}{4\pi\varepsilon_0|x-y|^2}\psi_1(x,y,t) \qquad\qquad\qquad\qquad\qquad (13.a) \\ i\hbar\dfrac{\partial}{\partial t}\psi_2(x,y,t) = -\dfrac{\hbar^2}{2m}\dfrac{\partial^2}{\partial x^2}\psi_2(x,y,t) + \left\{-\dfrac{\hbar^2}{2m'}\dfrac{\partial^2}{\partial y^2}\psi_2(x,y,t) + \phi(y)\psi_2(x,y,t)\right\} \\ \qquad\qquad + \dfrac{qq'}{4\pi\varepsilon_0|x-y|^2}\psi_2(x,y,t) \qquad\qquad\qquad\qquad\qquad (13.b) \end{cases}$$

Now, the potential difference $U(t)$ has disappeared from the eq.(13), i.e. the Schrodinger equations (13) are independent of the potential difference $U(t)$, so, the wave functions $\psi_1(x,y,t)$ and $\psi_2(x,y,t)$ should also be independent of $U(t)$. Comparing the equation (10) and (13), which are exactly same to each other, so, the following equations are obvious:

$$\begin{cases} \psi_1(x,y,t) = \psi_1^0(x,y,t) \\ \psi_2(x,y,t) = \psi_2^0(x,y,t) \end{cases} \qquad t_0 < t < t_3 \qquad (14)$$

In this situation, the total wave function is:

$$\begin{aligned}\psi(x,y,t) &= \psi_1(x,y,t) + \psi_2(x,y,t) \\ &= \psi_1^0(x,y,t) + \psi_2^0(x,y,t) = \psi^0(x,y,t)\end{aligned} \qquad t_0 < t < t_3 \qquad (15)$$

For $\psi(x,y,t)$ is the wave function of the charges $q$ and $q'$ with the potential difference between the two tubes being $U(t)$, but, $\psi^0(x,y,t)$ is the wave function with the potential difference between the two tubes being zero, $\psi(x,y,t) = \psi^0(x,y,t)$ means that the wave function will not change with the potential difference $U(t)$. So, when the two beams meet at F, the interference pattern will not change with $U(t)$, i.e. the phase shift as eq. (8) predicted by the A-B effect will not appear.

Why does not the phase shift predicted by the A-B effect appear if the induced charge is included?

Because, if there were no induced charge on the surfaces of the metal tubes, the electric field of the moving charge would exist in all the space, this field could penetrate the metal tubes and overlap with the electric field between the two tubes applied by the external voltage generator. While the moving charge $q$ lies in the tube 2, its potential energy is $qU(t)$; while the moving charge $q$ lies in the tube 1, its potential energy is 0. According to quantum mechanics, the wave function of the moving charge is:

$$\psi(x,t) = \psi_1(x,t) + \psi_2(x,t) = \varphi_1(x)e^{-iE_1 t/\hbar} + \varphi_2(x)e^{-iE_2 t/\hbar} \qquad (16)$$

where $E_1$ and $E_2$ are the energies of the parts passing through the tube 1 and 2, respectively. They are given by:

$$\begin{cases} E_1 = E_K \\ E_2 = E_K + qU(t) \end{cases} \qquad (17)$$

where $E_K$ is the kinetic energy of the moving charge. Obviously $E_1 \neq E_2$, so,

the time dependence of the $\psi_1(x,t)$ is different with that of $\psi_2(x,t)$. Therefore, when these two beams come out from the tubes, a phase shift due to the potential difference $U(t)$ between the tubes will appear, this is the electric A-B effect predicted by Aharonov and Bohm. This analysis is consistent with the eq.(8). The eq.(8) shows the phase shift $\Delta\varphi$ is proportion to $qV_0$, which represents the potential energy of the moving charge $q$. So, the eq.(8) strongly implys the phase shift $\Delta\varphi$ arise from the interaction energy between the moving charge $q$ and the electric field $U(t)$.

But in a real experiment, when the moving charge $q$ moves in the metal tubes, there must be an induced charge $q'$ appearing on the inner surfaces of the metal tubes. Just for the appearance of the induced charge $q'$, the total electric field produced by the moving charge $q$ and the induced charge $q'$ can only exist in the region enclosed by the inner surfaces of the metal tubes. So, this total electric field cannot overlap with the electric field between the tubes. The potential energy (related to $U(t)$) of the system including the moving charge $q$ and the induced charge $q'$ is zero no matter the moving charge $q$ lies in the tube 1 or tube 2. The wave function of the system:

$$\psi(x,y,t)=\psi_1(x,y,t)+\psi_2(x,y,t)=\varphi_1(x,y)e^{-iE_1 t/\hbar}+\varphi_2(x,y)e^{-iE_2 t/\hbar} \quad (18)$$

Where, $E_1$ is the energy of the system with the moving charge passing through the tube 1, $E_2$ is the energy of the system with the moving charge passing through the tube 2. $E_1$ and $E_2$ are given by:

$$\begin{cases} E_1 = E_K + E'_K + \phi(y) + \dfrac{qq'}{4\pi\varepsilon_0 |x-y|} \\ E_2 = E_K + E'_K + \phi(y) + \dfrac{qq'}{4\pi\varepsilon_0 |x-y|} + qU(t) + q'U(t) \end{cases} \quad (19)$$

where, $E_K$ and $E'_K$ are the kinetic energies of the moving charge $q$ and the induced charge $q'$; the other quantities are defined as above.

For $q=-q'$, so:

$$\begin{cases} E_1 = E_K + E_K^{'} + V(y) + \dfrac{qq'}{4\pi\varepsilon_0 |x-y|} \\ E_2 = E_K + E_K^{'} + V(y) + \dfrac{qq'}{4\pi\varepsilon_0 |x-y|} \end{cases} \quad (20)$$

Obviously,

$$E_1 = E_2 \quad (21)$$

So, the time dependence of $\psi_1(x,y,t)$ is same to that of $\psi_2(x,y,t)$. Therefore, when these two beams come out from the tubes, no phase shift due to the potential difference $U(t)$ will appear. When these two parts meet at F, the interference pattern will not change with the potential difference $U(t)$ between the two tubes.

In this situation, the potential difference $U(t)$ between the two metal tubes still exists, but the phase shift due to $U(t)$ does not appear. So, it is difficult to state the scale potential can influence the phase of a moving charge; it is more reasonable to state that the real factor influencing the phase of a moving charge should be the potential energy of the system including the moving charge $q$ and the induced charge $q'$.

**Acknowledgement**  The author thanks Mrs. Deng Chang-yu for her encouragement.  This work is supported by " the Fundamental Research Funds for the Central Universities, No: 2009JBM102 "

**CAPTIONS**

Fig. 1 , Schematic experiment to demonstrate the electric Aharonov-Bohm effect. $U(t)$ represents the external voltage generator.

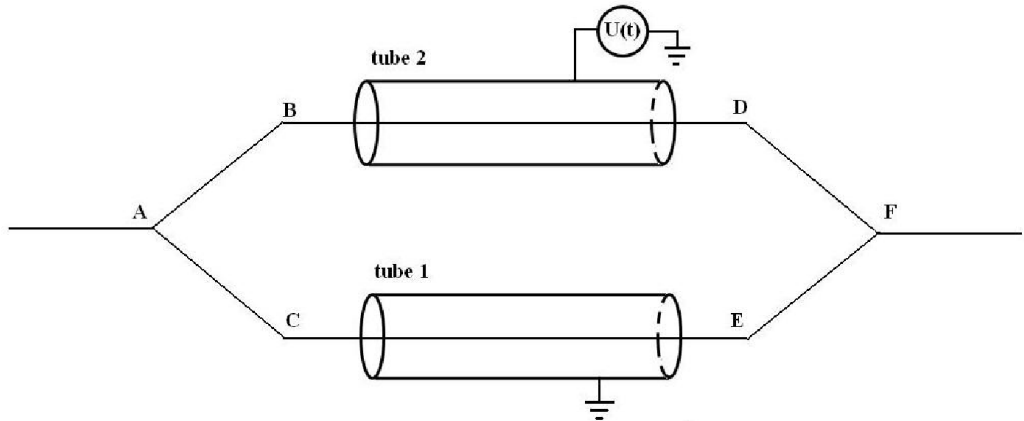

Fig. 1